# On the Origin of Tremor in Parkinson's Disease

Andrey Dovzhenok[1], Leonid L Rubchinsky[1,2]

[1] Department of Mathematical Sciences and Center for Mathematical Biosciences, Indiana University Purdue University Indianapolis, Indianapolis, Indiana, United States of America
[2] Stark Neurosciences Research Institute, Indiana University School of Medicine, Indianapolis, Indiana, United States of America

**Corresponding author:**
Leonid L Rubchinsky
Department of Mathematical Sciences
Indiana University Purdue University Indianapolis
402 N. Blackford St., Indianapolis, IN 46202, USA

E-mail: leo@math.iupui.edu;   Phone: 317-274-9745;   Fax: 317-274-3460

**Abstract**

The exact origin of tremor in Parkinson's disease remains unknown. We explain why the existing data converge on the basal ganglia-thalamo-cortical loop as a tremor generator and consider a conductance-based model of subthalamo-pallidal circuits embedded into a simplified representation of the basal ganglia-thalamo-cortical circuit to investigate the dynamics of this loop. We show how variation of the strength of dopamine-modulated connections in the basal ganglia-thalamo-cortical loop (representing the decreasing dopamine level in Parkinson's disease) leads to the occurrence of tremor-like burst firing. These tremor-like oscillations are suppressed when the connections are modulated back to represent a higher dopamine level (as it would be the case in dopaminergic therapy), as well as when the basal ganglia-thalamo-cortical loop is broken (as would be the case for ablative anti-parkinsonian surgeries). Thus, the proposed model provides an explanation for the basal ganglia-thalamo-cortical loop mechanism of tremor generation. The strengthening of the loop leads to tremor oscillations, while the weakening or disconnection of the loop suppresses them. The loop origin of parkinsonian tremor also suggests that new tremor-suppression therapies may have anatomical targets in different cortical and subcortical areas as long as they are within the basal ganglia-thalamo-cortical loop.

**Introduction**

Tremor is one of the cardinal symptoms of Parkinson's disease. Some studies report it to be present in up to 80% of patients with autopsy-proven Parkinson's disease [1]. It is a well-recognized feature of Parkinson's disease and is a disabling symptom. Parkinsonian tremor is primarily a rest tremor with the frequency in 3-7Hz range, it is episodic in time, can be modulated (suppressed or enhanced) by motor or cognitive activity; cortical and subcortical motor areas during episodes of Parkinsonian tremor exhibit bursty neuronal firing correlated with tremor EMG [2-5]. Tremor is believed to be different from akineto-rigid symptoms of the disease both in the patterns of degeneration of dopaminergic neurons [6] and in the spatial location and spectral content of the neuronal activity in the basal ganglia circuits [7-9].

While the occurrence of parkinsonian tremor is naturally related to dopaminergic degeneration (or, potentially, some other degeneration in Parkinson's disease), the network, cellular and synaptic mechanisms of parkinsonian tremor are not clear. It is commonly acknowledged that parkinsonian tremor has a central origin, but the localization of this central oscillator (oscillators) is still debatable. A few hypotheses of tremor generation have been proposed previously (reviewed in [4]). Some of them place an emphasis on the thalamus, suggesting that it either generates tremor because of the rebound activity of thalamic cells when they are released from excessive pallidal inhibition [10], or that it "filters" (converts) or otherwise promotes low-frequency oscillations out of a 10-15Hz frequency band [11,12]. Another suggests that the basal ganglia circuits may be the tremor-generating oscillator on its own [13]. However, the increase in interspike interval within the burst characteristic of the thalamic rebound bursting is not observed in thalamic bursts seen in parkinsonian tremor [14]. Neither is the thalamic filter hypothesis supported by data analysis [15] nor does it explain the origin of 10-15Hz oscillations. The tremor-suppressing effect of lesions outside the basal ganglia (such as lesions in the thalamus [2,14] and cortex [16]) suggests that the tremor generator may extend beyond the basal ganglia networks. Cerebellar circuits are involved in the tremulous movement, but appear to be not directly connected with the tremor movement [17] and are thus unlikely to be its generator (reviewed in [4]).

A very plausible view is that the tremor oscillator is localized in the basal ganglia-thalamo-cortical circuits (Figure 1A) (reviewed in [4]). The basal ganglia cells are known to possess rich membrane properties, which support pacemaking [18,19], but do not produce tremor oscillations in healthy basal ganglia circuits. In contrast, in parkinsonian circuits tremor-related activity (i.e. neural activity in the tremor frequency band, correlated with the tremor movement or tremor EMG) was observed in the basal ganglia (in the subthalamic nucleus, STN [20] and in pallidum [21]), in the thalamus [2,14], and in cortex [16,17]. Surgical lesions in different parts of the basal ganglia-thalamo-cortical loop (in the STN [22], in cortex (reviewed in [4]), in pallidum and the thalamus [23]) suppress tremor. The fact that breaking the loop at multiple sites leads to the same effect – tremor suppression – suggests that the loop itself, more than any of its parts, is a tremor generator. However, this evidence is indirect and does not tell how tremor is generated.

The present study explores the possibility of the basal ganglia-thalamo-cortical loop theory. We use computational neuroscience techniques to study the dynamics of this loop. While the physiology of the cortico-basal ganglia loops has been the subject of earlier computational studies (e.g., [24]), including studies [25,26], which provided further confirmation for the role of dopamine depletion in promoting oscillatory activity in various frequency bands (usually beta-band), these studies were mostly concerned with the action selection in basal ganglia and did not consider tremor oscillations. We show that the membrane properties of basal ganglia neurons together with anatomy of the basal ganglia circuits (Figure 1A) and the gross feedback-like structure of the basal ganglia-thalamo-cortical loop may generate tremor-like oscillations if the synaptic projections change their strength (as is expected to be the case in Parkinson's disease due to dopaminergic degeneration, discussed in more details in the Methods section). These tremor-like oscillations in the model are suppressed by breaking the loop in various locations. Deeper understanding of the tremor mechanisms will allow for improved treatment of parkinsonian tremor and will enhance our understanding of basal ganglia physiology.

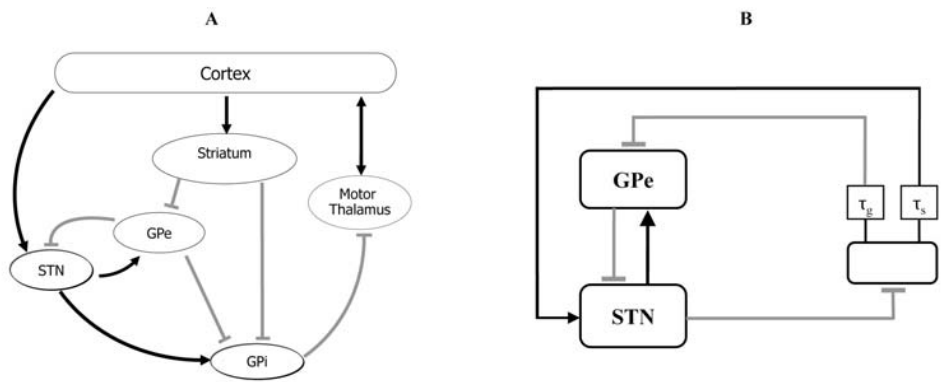

**Figure 1. Basal ganglia-thalamo-cortical circuit.** A) is the schematics of the anatomy. B) is the model circuit. There are one GPe and one STN neurons and a feedback neuron, represented by a feedback box. Arrows indicate excitatory synapses and bars indicate inhibitory synapses. Squares indicate the delay units with the delays $\tau_s$ and $\tau_g$.

## Methods

*Model circuit development*
*In vitro* studies [27] demonstrated how a cultured network of GPe and STN neurons can generate low-frequency oscillations. The bursting they observed is not necessarily the same as tremor oscillations. However, this experimental result indicates that pallido-subthalamic networks have the necessary cellular and synaptic properties to produce low-frequency oscillatory dynamics. Nevertheless, in healthy humans *in vivo*, subthalamo-pallidal networks do not generate tremor oscillations. Various studies (see Dopamine-dependent parameters subsection below) have provided evidence for how the basal ganglia-thalamo-cortical loop may become more strongly connected in Parkinson's disease. Lesions in different parts of the basal ganglia-thalamo-cortical loop in Parkinson's disease suppress tremor (see references in Introduction). This suggests an intriguing possibility that subthalamo-pallidal circuits embedded in a pathologically connected basal ganglia-thalamo-cortical loop may be the generator of tremor oscillations. Thus, the strength of the loop (defined by the underlying synaptic projections) becomes stronger in Parkinson's disease giving rise to tremor, while dopaminergic medication or surgical lesions would decrease the strength of or partially break the loop, suppressing tremor.

This reasoning suggests the following organization of the model network: a network of connected subthalamic and pallidal cells, embedded in a larger feedback loop (provided by basal ganglia, thalamic, cortical, and possibly other circuitry). To study the basic properties of the tremor oscillations in this loop, we consider a model, which retains some anatomical, synaptic and cellular properties of the underlying circuits, but simplifies others, especially those, which are not well-known. Thus, we suggest considering the circuit presented in Figure 1B. We retain the cellular properties of subthalamic and pallidal cells and circuits (which have some pacemaking properties [27]) by utilizing the detailed models of STN and GPe neurons developed by [28], but simplify the rest of the complex basal ganglia-thalamo-cortical network, which is represented in this study by a single neuron model and two delay units to incorporate the delays in the polysynaptic pathways the signals will travel through. This model network is in agreement with the known organization of the basal ganglia and related circuits [29], but obviously does not consider the detailed properties and parameters of the loop (which would be hard to estimate from experiments anyway). Thus, the results of the study will be sensitive to only general properties of the loop – essentially its presence or absence and overall connection strength.

The inhibitory input to the pallidal segment in the model represents thalamo-cortico-striatal and, possibly, thalamo-striatal [30] pathways. Excitatory input to STN represents the thalamo-cortico-STN pathway. GPi is not explicitly present in the model. Hence, the model architecture assumes GPi to be enslaved to STN input. While GPi intrinsic dynamics and non-STN inputs to GPi may affect the dynamics in the loop, the exclusion of GPi provides us with a model, which considers GPe-STN interaction with the feedback loop GPi projection to thalamus is inhibitory, which is reflected by inhibitory output of the model STN (in reality the latter sends excitatory projections to GPi).

*Model neurons*

Each module in the model circuit (Figure 1B) is represented by a single neuron modeled as a one-compartment conductance-based model. This is a gross simplification of a real system, where each module involves many neurons. Thus the present framework cannot describe effects such as, for example, convergence of synapses. However, tremor is a synchronized phenomenon and therefore combining many real neurons into one model neuron is reasonable. The neuronal models contain the experimentally observed conductances and thus the time scales of the real phenomenon. Hence, proposed modeling approach may provide a potential mechanism for tremorgenesis, although it does not exclude the existence of local network effects in real parkinsonian tremor. Relatively simple modeling components (cells and network) allow for a relative simplicity of model exploration and a low number of unknown parameters (such as synaptic strength and unknown connectivity patterns).

Since the properties of the subthalamic and pallidal cells are likely to contribute to the birth of oscillations (see previous subsection), we use conductance-based model of GPe and STN neurons developed in [28] on the basis of patch-clamp experiments and further utilized in an array of studies (e.g. [31-33]). On the contrary, the basal ganglia-thalamo-cortical feedback is represented by delay units and a very simple neuronal model. This generic model serves the mere purpose of the feedback signal propagation in the model circuit and does not include any further details of the loop architecture (see the reasoning above in the Model circuit development subsection). We also consider a more detailed version of thalamocortical relay model to study the robustness of the model feedback effect.

The models for GPe and STN neuronal modules (Figure 1B) include a leak current, fast spike-producing potassium and sodium currents, low threshold T-type and high-threshold $Ca^{2+}$-currents, and a $Ca^{2+}$-activated voltage-independent afterhyperpolarization $K^+$-current (AHP), so that the equation governing the membrane potential $V$ takes the form

$$C\frac{dV}{dt} = -I_L - I_K - I_{Na} - I_T - I_{Ca} - I_{AHP} - I_{syn} + I_{app}$$

with the membrane currents given as

$$I_L = g_L[V - V_L],$$
$$I_K = g_K n^4 [V - V_K],$$
$$I_{Na} = g_{Na} m_\infty^3(V) h [V - V_{Na}],$$
$$I_T = g_T a_\infty^3(V) r [V - V_{Ca}],$$
$$I_{Ca} = g_{Ca} s_\infty^2(V) [V - V_{Ca}],$$
$$I_{AHP} = g_{AHP} [[Ca]/[[Ca]+k_1]][V - V_K],$$

where $k_1$ is the dissociation constant of $Ca^{2+}$-dependent AHP current and square brackets denote multiplication. Note also that the effect of applied current $I_{app}$ in the membrane potential equations for the STN and GPe model neurons is to adjust resting membrane potentials to the experimentally measured values. It may be also achieved (perhaps in a more physiologically reasonable manner) by the adjustment of the leak current $I_L$ alone. However, we explicitly include $I_{app}$ for model uniformity with [28] and [32]. The intracellular calcium balance is described by the equation

$$d[Ca]/dt = \varepsilon[-I_{Ca} - I_T - k_{Ca}[Ca]],$$

where the constant $\varepsilon$ characterizes the calcium influx and the product $\varepsilon k_{Ca}$ is a calcium pump rate. The slow gating variables are described by the 1st order kinetic equation in the form:

$$dx/dt = \phi_x[x_\infty(V) - x]/\tau_x(V),$$

with time constant functions given by $\tau_x(V) = \tau_x^0 + \tau_x^1/[1 + \exp(-[V - \theta_x^\tau]/\sigma_x^\tau)]$, where $x$ can be $n$, $h$ or $r$. Note that the constant $\varphi$ in the equation for gating variable kinetics has no special meaning, but is left for uniformity with [28]. The steady state voltage dependent activation and inactivation functions for all gating variables have the form $x_\infty(V) = 1/[1 + \exp(-[V - \theta_x]/\sigma_x)]$, where $x$ can be $n$, $m$, $h$, $a$, $r$, or $s$. Here, $\theta_x$ is the half (in)activation voltage for the gating variable $x$ and $\sigma_x$ is its slope factor. The T current inactivation in the STN neuron is modeled with $b_\infty(r)$ instead of just variable $r$ as $b_\infty(r) = 1/[1 + \exp([r - \theta_b]/\sigma_b)] - 1/[1 + \exp(-\theta_b/\sigma_b)]$ following [28] for stronger rebound bursts. Voltage-dependent fast gating variables $m$, $a$ and $s$ are assumed to be instantaneous. GPe and STN neurons differ in parameter values (see Table 1), which were taken from [28], except for the parameter changes that follow [32] with applied current to STN further increased to $I_{app}$ =32 pA/μm² to produce more realistic firing rates.

The conductance-based model of the feedback neuronal module (Figure 1B) is a simple two-dimensional model, which includes the following equation for the membrane potential

$$C\frac{dV}{dt} = -I_L - I_K - I_{Na} - I_{S \to F} + I_{app}$$

with a simplified representation of a standard potassium spike-producing current (instantaneous activation, no inactivation). The potassium, sodium and leak currents are given by

$$I_K = g_K n[V - V_K],$$
$$I_{Na} = g_{Na} m_\infty(V)[V - V_{Na}],$$
$$I_L = g_L[V - V_L],$$

where square brackets denote multiplication. The gating variable $n$ has first-order dynamics

$$dn/dt = [n_\infty(V) - n]/\tau_n$$

with

$$n_\infty(V) = 1/[1 + \exp(-[V - \theta_n]/\sigma_n)]$$

while instantaneous activation of the Na⁺ current is given by

$$m_\infty(V) = 1/[1 + \exp(-[V - \theta_m]/\sigma_m)]$$

Mathematically, this model is similar to Morris-Lecar model. This neuron is tonically active (Figure 2C). The model and parameters (see Table 2) were taken from [34]. This is a very simple model of a neuron; this form is conditioned by a need to represent the thalamocortical feedback in a very simple and generic way. Two delay units were chosen to approximate the time it takes for the neuronal activity to travel through (potentially multiple) basal ganglia-thalamo-cortical loops before reaching STN and GPe regions. We use delay times $\tau_s$=30 ms and $\tau_g$=50 ms for the modeling reported below which appear to be physiologically plausible [35], however, eventually we explore a wide range of the delays.

We also considered the model circuit with the thalamocortical relay neuron [32] instead of the Morris-Lecar-type feedback neuron. Thalamocortical relay cell model current-balance equation has the form

$$C\frac{dV}{dt} = -I_L - I_{Na} - I_K - I_T - I_{S \to Th} + I_{ext}$$

where leak, sodium, potassium, and low-threshold calcium currents in the current-balance equation are

$$I_L = g_L[V - V_L],$$
$$I_{Na} = g_{Na}m_\infty^3(V)h[V - V_{Na}],$$
$$I_K = g_K[0.75[1-h]]^4[V - V_K],$$
$$I_T = g_T p_\infty^2(V)r[V - V_T],$$

with square brackets denoting multiplication.

The gating variables $h$ and $r$ have first-order kinetics governed by the equations $dx/dt = [x_\infty(V) - x]/\tau_x(V)$ ($x$ can be $h$ or $r$) where the voltage-dependent time constant functions are $\tau_h(V) = 1/[a(V) + b(V)]$ with $a(V) = \tau_a^1 \exp(-[V - \theta_a^\tau]/\sigma_a^\tau)$, $b(V) = \tau_b^1/[1 + \exp(-[V - \theta_b^\tau]/\sigma_b^\tau)]$ and $\tau_r(V) = \tau_r^0 + \tau_r^1 \exp(-[V - \theta_r^\tau]/\sigma_r^\tau)$. Steady-state activation and inactivation functions are $x_\infty(V) = 1/[1 + \exp(-[V - \theta_x]/\sigma_x)]$ (where $x$ can be $m$ or $p$) and $x_\infty(V) = 1/[1 + \exp([V - \theta_x]/\sigma_x)]$ (where $x$ can be $h$ or $r$). Parameters for this model (see Table 3) were taken from [71] except we set $I_{ext}$=0.85 that produced tonic firing activity with the frequency around 30Hz in the absence of other synaptic input.

Synaptic current $I_{syn}$ in the STN neuron is computed as a sum of synaptic currents from the GPe and feedback neurons: $I_{syn,STN} = g_{G \to S} s_G [V - V_{G \to S}] + g_{F \to S} s_F [V - V_{F \to S}]$. Similarly, synaptic current in the GPe neuron is a sum of currents from the STN and feedback neurons as $I_{syn,GPe} = g_{S \to G} s_S [V - V_{S \to G}] + g_{F \to G} s_F [V - V_{F \to G}]$. Synaptic current in the feedback neuron has the form $I_{S \to F} = g_{S \to F} s_S [V - V_{S \to F}]$. Here, the maximal synaptic conductance from neuron X to neuron Y is denoted $g_{X \to Y}$ and $s_X$ is a synaptic variable of the corresponding presynaptic neuron X, with X, Y taking values S, G and F for STN, GPe and feedback neurons, respectively. All connections in the model circuit are excitatory glutamatergic and inhibitory GABAergic synapses modeled by the 1$^{st}$-order kinetic equations describing the fraction of activated channels

$$ds/dt = \alpha H_\infty(V_{presyn} - \theta_g)[1 - s] - \beta s,$$

where $H_\infty(V) = 1/[1 + \exp(-[V - \theta_g^H]/\sigma_g^H)]$ is a sigmoidal function, $V_{presyn}$ is the membrane potential of a presynaptic neuron and the values of synaptic parameters $\alpha$, $\beta$ and $\theta_g$ in the STN and GPe neurons are taken from [28] (see Table 1), while feedback neuron excitatory (inhibitory) synaptic parameters were assumed to be the same as in STN (GPe) neuron. The values of synaptic strengths in the "normal" state (high dopamine level) are $g_{F \to G}$= 0.18, $g_{G \to S}$= 0.695, $g_{S \to F}$= 0.25, $g_{F \to S}$= 0.215, $g_{S \to G}$= 0.051, and the maximal conductance of the AHP current in STN neuron was set to $g_{AHP}$=4.23 nS/µm$^2$. The values of synaptic strengths corresponding to the parkinsonian (low dopamine level) state are $g_{F \to G}$= 0.36, $g_{G \to S}$= 1.39, $g_{S \to F}$= 0.5, $g_{F \to S}$= 0.43, $g_{S \to G}$= 0.103, with STN cell's AHP conductance set to $g_{AHP}$=8.46 nS/µm$^2$. Parameters for the parkinsonian state were found by varying synaptic strengths in physiologically relevant ranges to obtain distinct tremor-like activity in the model. Then, the normal state parameters were assumed to be 50% of their strength in the parkinsonian state. Further clarification of what a normal and a parkinsonian states are from the activity pattern standpoint are given in results section and Figure 2. Ultimately, we considered a large range of values for the synaptic strengths as we discuss below.

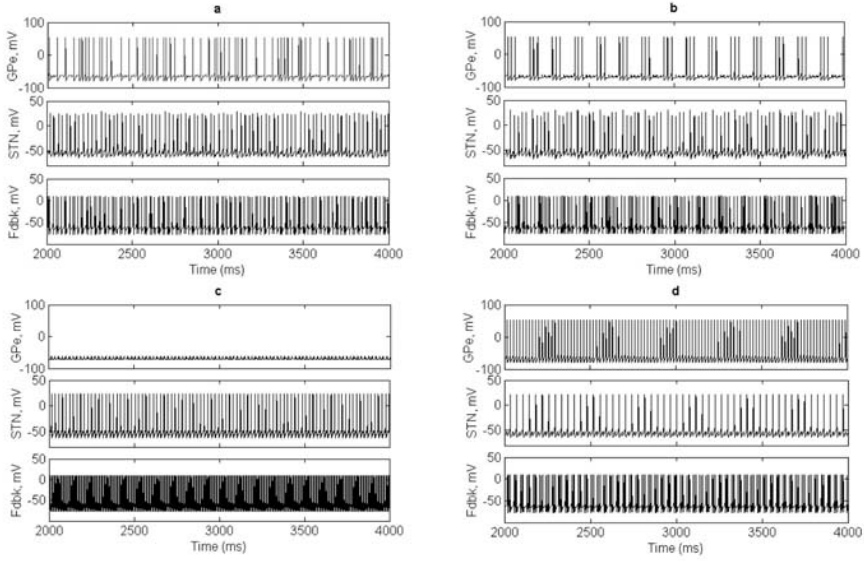

**Figure 2. Activity patterns in the model circuit in different states and after lesions.** A) "Normal" state, the activity pattern in GPe, STN and feedback neurons is tonic spiking. B) "Parkinsonian" state in the circuit with stronger feedback; STN and GPe neurons exhibit bursting discharge. C) the result of a lesion at the level of STN output, and D) the result of a lesion at the level of inputs to GPe and STN. While GPe is silent in C), STN firing is essentially tonic after both lesions, thus, the model basal ganglia circuit may generate tonic (and presumably more healthy) output. Parameter values in A) are: $s_1$=1.5, $s_2$=1.5. Parameters in B), C), and D) are: $s_1$=1, $s_2$=1.

The model circuit equations were simulated with XPP software (Bard Ermentrout, University of Pittsburg, http://www.math.pitt.edu/~bard/xpp/xpp.html).

*Dopamine-dependent parameters*
Because of the well-established dopaminergic degeneration in Parkinson's disease, the positive effect of L-DOPA on the symptoms (at least at the initial phase of treatment) and the tremor reduction produced by dopamine agonists [36], we study how the dynamics of the model system depends on parameters, which, in turn, are affected by the action of nigral dopamine.

Since nigral dopamine may modulate many types of synapses and cells in the basal ganglia, we consider two dopaminergic parameters, $s_1$ and $s_2$, which take into account several known dopaminergic actions. The first one, $s_1$, considers dopaminergic modulation of striato-pallidal and pallido-subthalamic synapses, and $s_2$ describes the modulation of cortico-subthalamic and subthalamo-pallidal synapses and of $Ca^{2+}$-activated $K^+$-current in STN. Dopamine is known to act on presynaptic receptors at striato-pallidal synapses reducing GABA release in GPe ([37]; see also [38]). In perhaps a similar manner, dopaminergic action in STN inhibits GABA release, in particular, from synapses from GPe [39-43]. These experiments also suggest that dopamine is able to suppress excitatory transmission to STN from cortex [44], while excitatory projections from STN to GPe are also suppressed by dopaminergic action [45]. Dopamine also has a tendency to depolarize STN cells by multiple mechanisms, in particular, including modulation of $Ca^{2+}$-activated $K^+$-current [46,47]. Overall, dopamine depletion seems to make the elements of the basal ganglia circuitry more functionally connected (e.g., [19]).

Thus, we set up two dopaminergic parameters $s_1$ and $s_2$ to modulate synaptic or membrane conductance to make them weaker or stronger, as one expects them to be in the presence or absence of dopamine. A dopamine-modulated conductance $g = (2 - s_i)g_0, i = 1,2$, where $g_0$ for $s_1$ involves $g_{F \to G}$ and $g_{G \to S}$, and $g_0$ for $s_2$ involves $g_{F \to S}$, $g_{S \to G}$, and $g_{AHP}$. We usually vary $s_1$ and $s_2$ in the [1, 2] range, so that lower values of $s_1$ and $s_2$ correspond to lower dopamine levels and stronger conductances. As dopaminergic parameters $s_{1,2}$ are decreased from 2 to 1, conductance $g$ increases from 0 to some maximal value $g_0$ which would correspond to the transition from high to low dopamine level (with transition from normal to parkinsonian state presumably being somewhere within these bounds). Obviously, the real modulation by the dopamine may not necessarily scale in the same way for all of its targets and is unlikely to go all the way to 0. Above division of parameters affected by dopamine into two groups is somewhat arbitrary. But our approach allows us to explore the parameter space in the model when $s_1$ and $s_2$ are changing in a particular direction (of increasing or decreasing dopamine level) over a large range. Exploration of the two-parametric space is a compromise, which allows us to avoid exploration of a high-dimensional parametric space (which may be hard to interpret anyway), yet, lets us study what happens with the network dynamics, when more than just one dopamine-sensitive parameter is being modulated.

*Time-series analysis*
To quantify the presence of tremor-like oscillations in the modeling circuit we used a modified version of the signal to noise ratio (SNR) criterion adopted in [48] to study the dynamics of tremor in parkinsonian patients:

$$SNR1 = \frac{\max_{\omega_a \leq \omega \leq \omega_b} \{P(\omega)\}}{\text{avg}_{\omega_{\min} \leq \omega \leq \omega_{\max}} \{P(\omega)\}}$$

where $P(\omega)$ is the power spectrum. This SNR criterion is used here to measure the degree of bursting activity in the tremor frequency range in the STN model neuron. The parameter setting of this criterion are $\omega_a$=4Hz, $\omega_b$=8Hz for the tremor band $[\omega_a, \omega_b]$ and $\omega_{\min}$=3Hz, $\omega_{\max}$=30Hz for the wider band $[\omega_{\min}, \omega_{\max}]$. While the real parkinsonian tremor may present with frequencies slightly lower than 4Hz, the 4-8Hz range in the model appears to be sufficient to study the bursting in the system. Moreover, proprioceptive feedback tends to lower parkinsonian tremor frequency [49,50]. This sensory feedback is not a part of the central mechanisms represented by the model.

To avoid analyzing transients, we ran simulations for 3s first and used the next 8.2s for time-series analysis. The time-series analysis steps were similar to those in [48]. The time-series of STN voltage was cut into non-overlapping intervals of equal length of around 0.8s, multiplied with a Hanning tapering window and processed with fast Fourier transform (FFT) for each interval in the data sample. Obtained values were normalized by the interval size. Finally, SNR was calculated as a mean of values for each time interval. Only time-averaged SNR was considered in the current paper.

To show the robustness of tremor detection we introduced three more variations in SNR criteria. The second SNR criterion identified the position $\omega_m$ of the peak of the power spectrum in 3-8 Hz range to create a $\Delta\omega$ frequency band centered around this peak: $[\omega_m - \Delta\omega/2, \omega_m + \Delta\omega/2]$ and then computed

$$SNR2 = \frac{\max_{\omega_m - \Delta\omega/2 \leq \omega \leq \omega_m + \Delta\omega/2} \{P(\omega)\}}{\text{avg}_{\omega_{\min} \leq \omega \leq \omega_{\max}} \{P(\omega)\}}$$

with $\Delta\omega = 4Hz$. Therefore, SNR2 was supposed to identify oscillations in a band of the same width, but different center frequency than SNR1 (such as 3-7Hz, 5-9Hz, etc.), detecting oscillations in part of the spectrum slightly wider than usual parkinsonian frequencies.

The other two criteria used average power in the fixed 4-8Hz tremor band or average power in the floating band around the peak $[\omega_m - \Delta\omega/2, \omega_m + \Delta\omega/2]$, instead of the maximal values, i.e.

$$SNR3 = \frac{\underset{\omega_a \leq \omega \leq \omega_b}{avg} \{P(\omega)\}}{\underset{\omega_{min} \leq \omega \leq \omega_{max}}{avg} \{P(\omega)\}}$$

and

$$SNR4 = \frac{\underset{\omega_m - \Delta\omega/2 \leq \omega \leq \omega_m + \Delta\omega/2}{avg} \{P(\omega)\}}{\underset{\omega_{min} \leq \omega \leq \omega_{max}}{avg} \{P(\omega)\}}$$

While not completely equivalent, these criteria are similar, as intended, since they all are aimed at identification of oscillations. All time-series analysis was performed in MATLAB (MathWorks, Natick, MA).

## Results

*Tremor oscillations in the model of basal ganglia-thalamo-cortical loop*
Although pallidal and subthalamic cells and their computational models used here are known to possess burst properties (see Introduction) under certain conditions, the modeling network (see Figure 1B) exhibits tonic spiking activity under moderate values of the coupling strength ($s_1$=1.5, $s_2$=1.5) (Figure 2A). We consider these dynamics as the normal (healthy) state, as no tremor-like oscillations are present in the modeling circuits.

As the coupling increases ($s_1$=1, $s_2$=1), STN and GPe neurons in the model network exhibit bursting activity (Figure 2B) with a frequency around 6Hz. This kind of dynamics, with bursting in the STN neuron at the tremor range is considered here as a parkinsonian state, because it exhibits tremor-like oscillations.

To further explore the relevance of these model oscillations to the real tremor we will study the dynamics of the model in response to the modifications of the network, representing dopaminergic treatment (see next subsection) and therapeutic lesions used to suppress tremor. There is no explicit representation of GPi in the model network, so that pallidotomy may be represented in the model by removing the projection from STN to the thalamo-cortical circuits. When this projection is removed from the model in the parkinsonian state ($s_1$=1, $s_2$=1) the STN activity is almost tonic (this will be quantified in the next subsection with the SNR criteria, as described in Methods). Even though GPe is silent here (presumably due to stronger inhibition from the feedback neuron in the absence of subthalamic inhibitory input), the tonic nature of STN discharge (Figure 2C) confirms that the system returns in a normal state.

The other kind of lesion reproduced in the model is at the level of cortex or thalamus. In the model that would correspond to a lesion of inputs to the GPe and STN segments (or, in other words, removal of the feedback neuron). In this case, the activity patterns of both GPe and STN are switched to tonic firing (Figure 2D). Note that in the case of this lesion at the level of basal ganglia input the feedback neuron shows somewhat bursty output. However, this bursting activity is at a much higher frequency and, therefore, cannot lead to tremulous movement of limbs. The characteristic feature of GPe neuron is its tonic activity and high firing rate.

While the interpretation of the model lesions is not unique (and is left for Discussion), in both lesion cases the feedback is removed in one way or another and in the "normal" case the feedback is weakened. The next subsection provides a systematic study of the circuit behavior for varying feedback.

*The effect of dopaminergic modulation*

To study the effects of dopaminergic modulation we varied dopaminergic parameters $s_1$ and $s_2$ as proxy for the presence of dopaminergic modulation (see Dopamine-dependent parameters subsection of Methods). The results of the previous subsection suggest that the strength of the feedback is essential for the occurrence of bursting, so we varied the dopaminergic parameters in a broad range to see how bursty the discharge is (as quantified by SNR criterion, see Methods).

As an example, we consider the SNR1 as we vary the dopaminergic parameter $s_1$ in the interval [1,2] (Figure 3). As the dopaminergic parameter increases, SNR1, which indicates the presence of the tremor-related bursting (the presence of oscillations in the tremor frequency band), decreases, first moderately, then sharply to less than 1 (the lack of activity in the tremor frequency range). Thus, Figure 3 illustrates the transition between tremulous and non-tremulous case, as the dopaminergic action changes. Of note is a relatively sharp onset of tremor oscillations in the model and jagged profile of SNR. We think this is most likely due to the simplicity of the model. While gross structure (strong coupling – oscillations; weak coupling – no oscillations) is captured by the model, the exact details of oscillatory/nonoscillatory transition in the model depend on a particular set of bifurcations the model experiences as the parameters are varied. This bifurcation cascade is likely to be model-specific. Moreover, if dopamine-dependent parameters are varied in different ways, the SNR profile may be different.

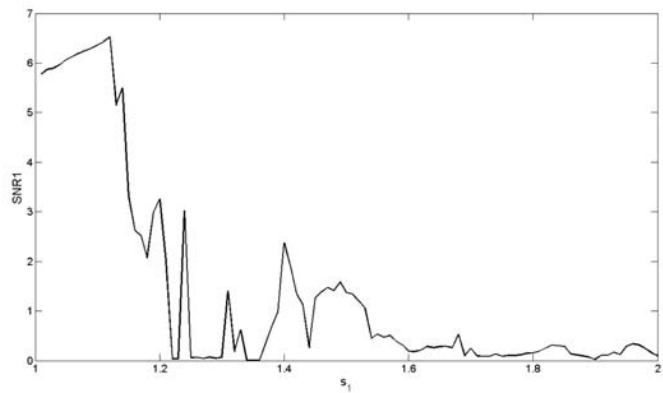

**Figure 3. The presence of tremor frequency band activity in the STN neuron as measured by the value of SNR1 in dependence on the dopaminergic parameter $s_1$.** The dopaminergic parameter $s_2$=1.1.

While the example above may be illustrative of the role of the dopamine-modulated thalamo-cortical feedback loop, the results of dopamine action on different synapses and cells in the system may be different. As we explain in the Methods, we study the effect of independent modulation of different properties of the network employing two dopaminergic parameters $s_1$ and $s_2$. How exactly dopamine will affect different synaptic and cellular parameters is not known, but the independent variation of two dopaminergic parameters (which, in turn, corresponds to variation of several synaptic and cellular parameters, see Methods) should give some general knowledge about the effect of the basal ganglia-thalamo-cortical feedback loop on the tremor-like bursting in the basal ganglia circuits.

We varied both $s_1$ and $s_2$ in the range from 1 to 1.9, which corresponds to the variation of the underlying network parameters from some maximal values to almost zero. The presence of tremor-like activity in STN (the output node of our simplified basal ganglia network) was assessed with SNR criteria. Figure 4 presents the result of this numerical experiment. Four different sub-plots were generated with the different SNR criteria (SNR1-4). Lighter shade of grey indicates stronger tremor activity. It is hard to define the exact level of SNR above which the activity can be called tremulous. However, the present SNR criteria are based on an earlier experimental study of tremor in parkinsonian patients [48], which uses 3.7 as a critical value for SNR1. The four SNR criteria employed here are slightly different one from another and the resulting subplots in Figure 4 are also slightly different. In particular, maximal SNR tends to yield larger values than those of averaged SNR, which may be attributed to the large height and small width of the spectral peaks. However, overall, tremor-like activity is present in the same regions, regardless of the criteria used. The smallness of the differences between the subplots points to the generic character of the observed phenomena.

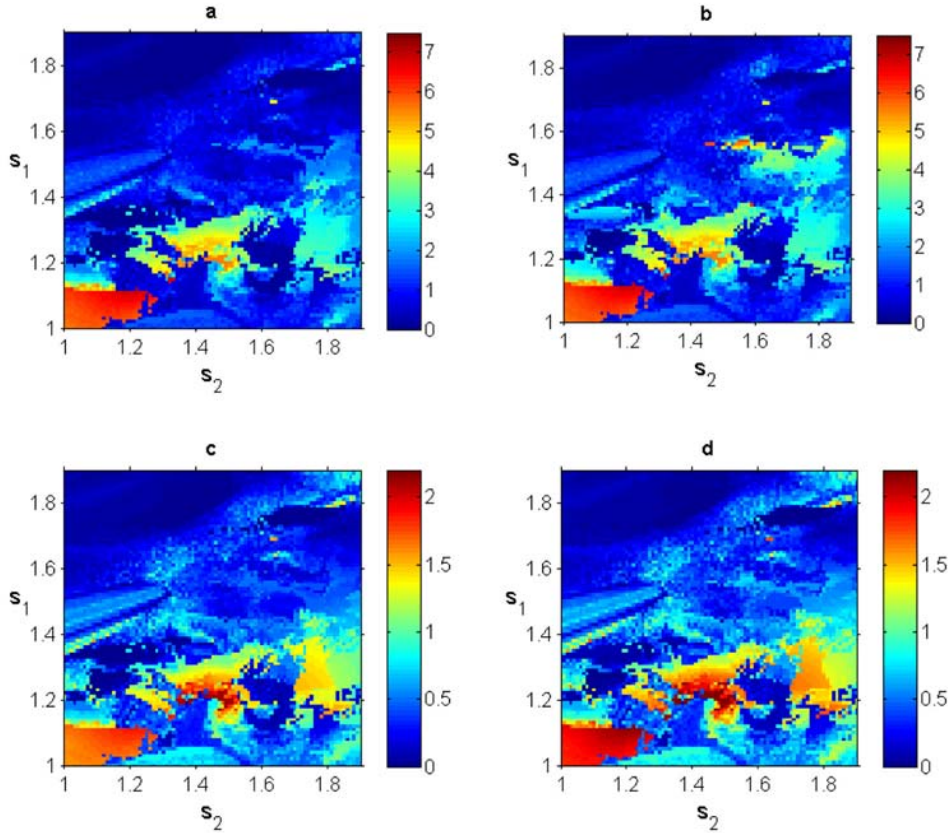

**Figure 4. Tremulous activity with variation of dopaminergic parameters.** The parameters $s_1$ and $s_2$ run along vertical and horizontal axis respectively, color codes for the value of SNR. The point (1, 1) corresponds to the bursting mode shown in Figure 2B. A), B), C), and D) represent SNR1, 2, 3, and 4 respectively. Parameters $s_1$ and $s_2$ are proxies of dopaminergic status and their higher values correspond to stronger dopamine influence. Thus upper right corner corresponds to a "normal" state of the network, while lower left corner corresponds to a "parkinsonian" state. Blue color indicates the absence of tremor-band oscillations, red color indicates prominent oscillations. Yellow and green correspond to SNR values termed to be tremulous in [48].

Figure 4 shows that the low values of dopaminergic parameters, i.e. low $s_1$ and $s_2$, tend to promote bursting in the tremor frequency range. This indicates that the strength of the coupling in the basal ganglia-thalamo-cortical feedback loop is responsible for the tremor oscillations. However, the dependence of SNR on the dopaminergic parameters is not monotonic. The areas of high SNR are interspersed with the areas of low SNR. The relative contribution of $s_1$ and $s_2$ is also different. Nevertheless, the general pattern (low dopaminergic parameter values – more tremulous activity, high values – less tremulous activity) is persistent.

*The effect of calcium, AHP and T-type currents*

We also study the effect of individual currents on the tremor-like oscillations in the loop. Figure 5A shows how the dopaminergic action parameter $s_1$ and the conductance of the AHP current in STN neuron $g_{AHP}$ affect SNR of tremor frequency oscillations. Tremor-like oscillations exist in a region of relatively strong values of the AHP current conductance and the parameter $s_1$ around the parkinsonian state of Figure 2B ($s_1=1$, $g_{AHP}=1$). Unlike Figure 4, dependence of SNR on the dopaminergic parameter $s_1$ and the AHP current conductance alone is monotonic: SNR abruptly decreases and remains low indicating disappearance of bursting activity in the tremor band as the value of $g_{AHP}$ decreases. Hence, the tremor-like oscillations in the loop substantially depend on the AHP current.

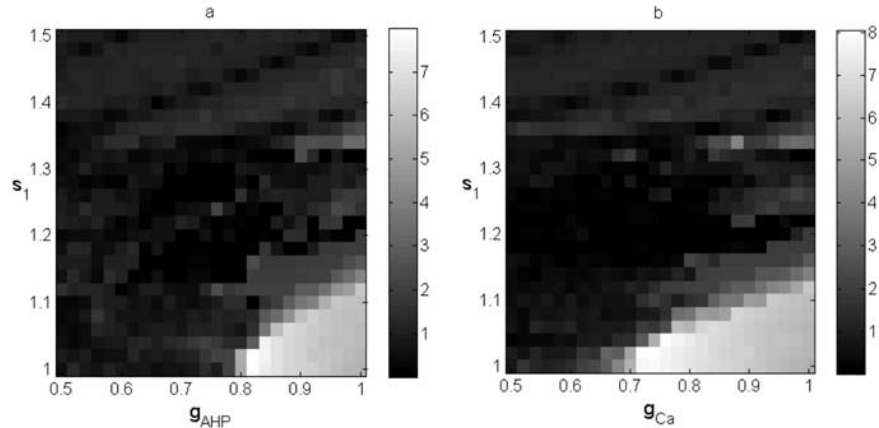

**Figure 5. Tremulous activity with variation of current conductances.** The shades of gray code for the value of SNR1 (like the color code in Figure 4) so that lighter areas exhibit stronger tremor oscillations. A) Bursting activity with variation of the dopaminergic parameter $s_1$ and the AHP current conductance $g_{AHP}$. B) Bursting activity with variation of the dopaminergic parameter $s_1$ and the Ca current conductance $g_{Ca}$. Parameters are the same as in Figure 2B. $g_{AHP}$ and $g_{Ca}$ are in units of the AHP current and the Ca current conductances in Parkinsonian state respectively ($g_{AHP}=8.46$, $g_{Ca}=0.5$). Lower values of AHP and Ca currents conductances lead to disappearance of tremor.

Similar results are obtained when we varied the $Ca^{2+}$ current conductance together with the dopaminergic parameter $s_1$ as shown in Figure 5B. Again, the dependence of SNR on the parameters is monotonic: tremor-like oscillations in the model exist in a single region around ($s_1=1$, $g_{AHP}=1$) and disappear when the value of the $Ca^{2+}$ current conductance is lowered. This similarity may be due to the fact that the decrease in the $Ca^{2+}$ current lowers the intracellular $Ca^{2+}$ concentration and therefore leads to reduction in the calcium-dependent AHP current in STN neuron.

Interestingly, our study revealed no substantial dependence of oscillations on the T-type current (not shown). In the model circuit STN neuron's T-type current is almost inactivated and cannot deinactivate due to relatively small inhibition from the GPe neuron. Hence, these results may indicate that the T current is not strongly involved in generation of tremor-like activity.

*The role of slow calcium dynamics in tremor-like bursting*
Next, we investigated the effect of calcium dynamics on tremor-like bursting activity in the model network. Figure 6 shows the presence of tremor-like burst firing in the STN cell as measured by SNR1 with varying $Ca^{2+}$ dynamics in the STN and GPe neurons. We made calcium dynamics artificially slower or faster by multiplying calcium constant $\varepsilon$ (in both STN and GPe neurons) by coefficients $\varepsilon_s$ and $\varepsilon_g$. First we slowed down the calcium dynamics as shown in Figure 6A. Here, sustained tremor-like oscillations exist in the STN unit until $Ca^{2+}$ dynamics in the GPe neuron becomes almost an order of magnitude slower ($\varepsilon_g \approx 0.2$). Around this value, bursting activity in the tremor frequency range changes to tonic firing and this transition is mostly independent from calcium dynamics in the STN cell as can be seen by a nearly vertical transition from tremor oscillations (red and yellow) to tonic firing (blue) (Figure 6A). Similar results are obtained when calcium dynamics is accelerated (Figure 6B). Tremor-like oscillations are maintained in the STN neuron until $Ca^{2+}$ dynamics becomes around an order of magnitude faster at which point regions of tonic firing activity in the STN cell become interspersed with bursting in the tremor frequency range (Figure 6B). Similarly to Figure 6A, independence from calcium dynamics in STN neuron is mostly maintained. Interestingly, in both cases (Figure 6A and 6B), the change in $Ca^{2+}$ dynamics did not affect tremor frequency which was almost constant at around 7.5Hz up until the transition to tonic firing activity (not shown). These results strongly suggest that the tremor-like oscillations in the STN neuron have network origin and are not solely based on the time scale of calcium dynamics in the STN and GPe neurons.

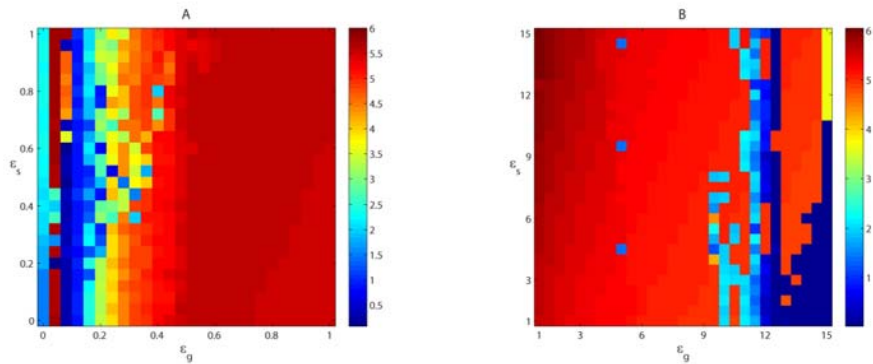

**Figure 6. Tremulous activity with the change in slow calcium dynamics.** A) Bursting activity with slowing calcium dynamics in the STN and GPe neurons. B) Bursting activity with accelerating calcium dynamics in the STN and GPe neurons. Color codes for the value of SNR1 (similar to Figure 4) so that yellow and red areas represent tremor oscillations. The parameters are as in Figure 2B. $\varepsilon_s$ and $\varepsilon_g$ are dimensionless constants to slow down or speed up the calcium dynamics in the STN and GPe neurons, correspondingly (see text). Tremor oscillations are robust with significant variations of time scale of calcium dynamics in STN and GPe.

*The influence of delays in the basal ganglia-thalamo-cortical loop*
The model circuit (Figure 1B) incorporates two delay units, which represent synaptic and conductance delays in polysynaptic pathways from STN to GPi to thalamus to cortico-striatal system. While these delays are likely to be fixed for each individual subject, we do not know their exact values. Therefore we study the impact of the delays on the tremor-like activity in the model network. Both delays, $\tau_s$ and $\tau_g$, were varied independently in a relatively large range. This range may include biologically unrealistic delay values but the objective is to ensure that the real delays are in the domain studied.

Figure 7 describes how delays affect SNR of tremor frequency oscillations. The regions of tremulous activity in the plane of delays are in the form of relatively narrow stripes; the slope of these stripes does not vary much and is close to 1. This suggests that the difference between delays may be more important than the values of the delays. Figure 7 also indicates that the oscillations are robust with respect to variation in delay values and tremor-like bursting exists for multiple values of the delays. Thus even though the exact values of the delays in the loop are not known, there are likely to be some fitting with those at the domains of tremor existence.

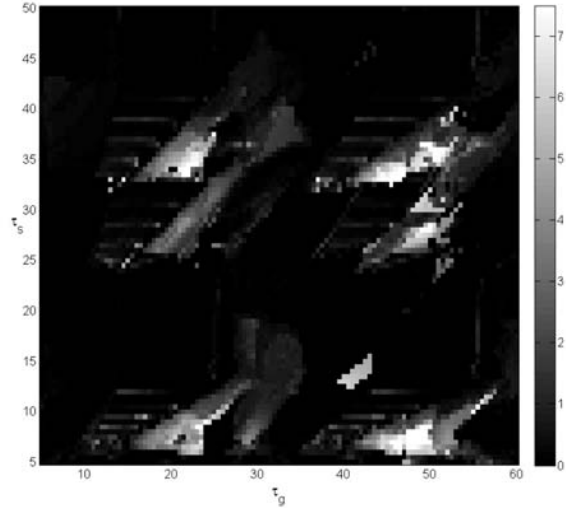

**Figure 7. Tremulous activity with variation in delays.** The parameters are the same as in Figure 2B. The shades of gray code for the value of SNR1 (like the color code in Figure 4) so that lighter areas exhibit stronger tremor oscillations (grey and white areas are tremulous dynamics).

*The influence of the feedback neuron model on the dynamics of the network*
Finally, we substitute the Morris-Lecar-type feedback neuron considered so far in this paper with a more physiologically realistic thalamocortical relay cell model in the form used in [32], which includes sodium, potassium and leak currents, as well as low-threshold calcium current. Figure 8 shows how SNR depends on the dopaminergic parameters $s_{1,2}$ in the case of this modified model circuit. Similarly to the simple feedback model (Figure 4), STN oscillations in tremor frequency-band exist in the model network in parkinsonian state and cease when the dopaminergic parameter $s_1$ increases to indicate higher dopamine level and presumably normal state. This suggests that the observed dynamics are robust with respect to different types of the feedback neuron in the model network. In turn, this suggests that the delayed feedback loop itself is likely to be essential for tremor-like oscillations in the model together with the cellular properties of STN and GPe neurons.

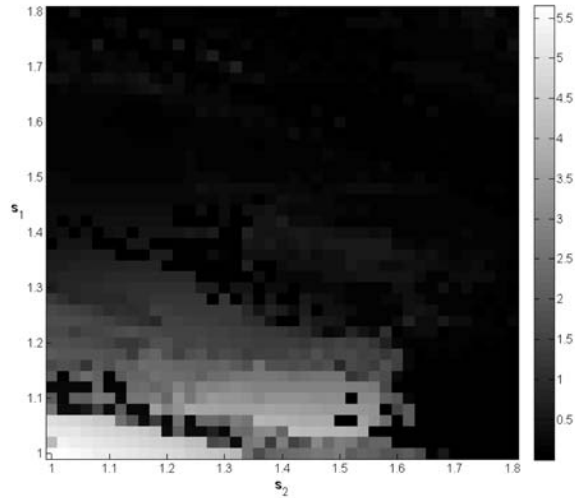

**Figure 8. Tremulous activity with a thalamocortical relay cell instead of a feedback neuron in the thalamocortical feedback loop.** The point (1, 1) corresponds to parameter values as in Figure 2B, except $g_{STN \to Th}$=0.46. Thalamocortical neuron parameters are given in Table 3. The parameters $s_1$ and $s_2$ run along vertical and horizontal axis respectively, the shade of grey codes for the value of SNR1.

## Discussion

*Summary of the modeling results*
The modeling shows that anatomical and membrane properties of subthalamo-pallidal circuits are prone to generation of tremor-like bursting in the presence of relatively strong basal ganglia-thalamo-cortical feedback. As we strengthen synaptic projections in the network (the expected outcome from the lack of dopaminergic modulation in Parkinson's disease), the tremor-like oscillations become more prominent. The destruction of the feedback leads to the suppression of the tremor-like oscillations (as one would expect from the outcomes of surgical lesions in parkinsonian patients).

The dependence of the strength of tremor-like oscillations on the strength of dopamine-dependent synaptic projections is not monotonic. Based on the simple model setup, one can hardly specify which range of synaptic parameters corresponds to the actual range of variation of the synaptic strength experienced in Parkinson's disease. Moreover, the effects of adding in dopamine agonist (which guided the choice of parameters modulated by dopamine) are not necessarily opposite to the effects of dopamine depletion taking place in Parkinson's disease. However, the model study demonstrates the general pattern of the change: as the basal ganglia-thalamo-cortical feedback loop becomes stronger, oscillations are likely to occur. The phenomenon is robust with respect to different kinds of modulation of the dopamine-dependent parameters. The phenomenon is also robust with respect to different values of delays in the feedback loop. While the actual delays are not likely to change in Parkinson's disease, they are not well-known. But the studied phenomenon persists for different values of delays.

Interestingly, recent studies suggest that the dopamine depletion negatively impacts autonomous activity in GPe [51]. Such a decline in GPe pacemaking may be seen at least to some extent similar to the increase in synaptic coupling, since in both cases the degree to which intrinsic dynamics influences the overall activity of a neuron is diminished in comparison with synaptic influence.

There are two important observations regarding the properties and mechanism of tremor oscillations in the model.

We simulated the dependence of the tremor-like oscillations on the time scale of calcium dynamics in STN and GPe, and on the strength of calcium and calcium-dependent potassium current. The calcium and calcium-dependent potassium currents need to be sufficiently strongly expressed to yield tremor-like oscillations. However, interestingly enough, the presence of tremor-like oscillations (and even their frequency) is stable over a large range of the time-scale of slow calcium dynamics (time scale of calcium $\varepsilon$ in both GPe and STN cells). Neither does it rely on the presence of T-type calcium current (which is almost inactivated during tremor-like oscillations in numerical experiment). This suggests that disruption of high-threshold calcium current and calcium-dependent potassium current, but not calcium time scale or T-type calcium current, will affect the existence of tremor. Eventually this may be an interesting statement to test experimentally.

Numerical simulation also indicates the importance of delays in the thalamo-cortical feedback. However, the delays in the circuitry are much shorter than the period of oscillations. This again suggests the importance of the network effects and of cortico-subcortical interactions for the genesis of tremor and setting its frequency. Delays are hard to manipulate with experimentally; however, from a theoretical standpoint, it will be very interesting to study how the delays may promote oscillations of a much longer period.

Finally we would like to note that the model effectively utilizes a negative delayed feedback (just follow the signs of synaptic connections in the loop for the STN unit, Figure 1B), which is known to be able to give rise to oscillations [52]. A generic model for parkinsonian tremor with delayed negative feedback was studied by [53]. Their model, however, was concerned with delayed proprioceptive feedback which had long been shown not to be significantly involved in the origin of parkinsonian tremor [49,50,54,55]. It also did not represent the cortico-subcortical circuitry and membrane properties of the involved cells. In that respect it was a more generic study of how the feedback may influence oscillations. In the current study we consider the cellular models with appropriate membrane properties, realistic network anatomy, the modulation of the network due to the lack of dopamine, and the results of known surgical interventions in Parkinson's disease and how they affect tremulous activity. We consider the feedback mechanism in Parkinsonian context. Thus this study provided computational rationale to suggest that this is the dopamine-mediated strength of cortico-subcortical loop, which facilitates the birth of tremulous oscillations.

The very general nature of the feedback in the model and the robustness of the studied phenomenon indicate that the details of the feedback are unlikely to produce a substantial qualitative change in the modeling results.

*Limitations of the modeling*
The model considered clearly has some limitations. The simplicity of the model basal ganglia-thalamo-cortical feedback is both its advantage (as it provides a way to study the generic effects of the feedback) and disadvantage (as it limits the model in many ways). Several limitations are discussed below.

The model network includes only single STN and GPe neurons following the framework of minimalistic approach to modeling. There are two different ways, in which this may limit the conclusions of the study. First is the very limited representation of the circuitry. The real anatomy of cortico-subcortical loops is complex while we consider simplified representation of striatum, thalamus and cortex and omit the other brain structures related to cortico-subcortical motor circuits. The minimal circuit considered naturally cannot tell anything about particular effect of this anatomy; however, it suggests that the observed phenomenon is robust, may be generated due to the feedback as a general anatomical feature and may be not very sensitive to the details of the circuitry.

Second is the modeling of a whole nucleus with a single neuron. As it was discussed in the modeling subsection this neglects a potential synaptic convergence and associated effects. So, what is considered here is essentially the case of synchronized oscillations, which appears to be a reasonable case for tremor. In addition, the complexity of the real network and the number of possible (and sometimes unknown) connection parameters in the loop is huge. The introduction of these elements into the model will substantially increase the number of unknown parameters. In particular, earlier modeling studies [28] considered oscillations within the basal ganglia network (but not cortico-subcortical loops) and the effect of the intrapallidal connectivity on these oscillations. However, the oscillations considered in that framework (which are likely to correspond to the beta-band oscillations accompanying hypokinetic symptoms of Parkinson's disease) are not related to tremor oscillations and can be supported by networks without intrapallidal connectivity [33,56].

While overall the model neurons exhibit reasonable patterns of neural activity, the case of the model lesions at the level of basal ganglia output may present some problem. When the STN firing does not exhibit tremulous activity, GPe is silent (Figure 2C). Hence, our modeling predicts the reduction in GPe activity after GPi lesion. It is hard to know if this is what happens in parkinsonian patients after lesions in internal pallidum (we are not aware of recordings in GPe after lesion in ipsilateral GPi in patients with tremor). Moreover, some studies indicate that GP intrinsic oscillation capability may increase after STN lesion [57] which would break the feedback to the basal ganglia. However, STN lesion also removes direct STN input to GPe and thus is not equivalent to GPi lesion in our model. Thus we believe the numerical studies still indicate that a stronger basal ganglia-thalamo-cortical feedback promotes tremor oscillations and its destruction suppresses them. Even though the lack of activity is visible in GPe in the computational results, the output of the model basal ganglia lacks burstiness. Likewise, in the case of the model lesions at the level of basal ganglia input the feedback neuron shows bursty output. Nevertheless, this bursting activity is high-frequency and, therefore, cannot give rise to tremor. Given the minimalistic modeling approach, the model should not be expected to reproduce the results of all known cortical and subcortical lesions with high fidelity. Similarly, minimalistic modeling approach may not necessarily reproduce the firing rates with high fidelity.

The traditional target for anti-tremor thalamotomy in Parkinson's disease is thalamic nucleus ventralis intermedius, Vim, although basal ganglia projections to thalamus target the nucleus ventro-oralis posterior, Vop [58]. Vim is not directly represented in our model circuit. However, Vim is effective site for surgical treatment (whether lesion or deep brain stimulation) of many tremor types beyond Parkinson's disease [59,60]. This does not necessarily indicate that Vim is the ultimate tremor-generator, rather it may be a downstream part of the circuitry between tremor generator and limbs. Thalamus may be a "bottleneck" for cortico-subcortical circuits involved in tremor generation and maintenance [61]. As we discussed in the introduction, cerebellar networks, while involved with the parkinsonian tremor movement, are unlikely to generate it directly [4,17]. Moreover, the basal ganglia (in particular, STN) have a disynaptic projection to the cerebellar cortex, which can be a way for basal ganglia dynamics to affect the activity in cerebellar circuits [62].

We did not consider the effect of deep brain stimulation (DBS) on tremor in the model. DBS may have differential effects on various neuronal elements, which are not present in the model (e.g., [63]). Nevertheless, the complicated network effect of DBS appears to perform "informational lesion", i.e. functionally disrupt the flow of pathological signals through the basal ganglia-thalamo-cortical loop (see, e.g., [64-66]). Thus the effect of DBS in the context of the present minimal model may be equivalent to that of a lesion.

The dopaminergic system is not the only transmission and modulation system affected in Parkinson's disease. Cholinergic and serotonergic disruptions have been observed as well (discussed in e.g., [7]).

Tremor severity in Parkinson's disease is poorly correlated with the degree of dopaminergic denervation, at least in striatum. Nevertheless, even in cases of Parkinson's disease with tremor only, i.e. monosymptomatic rest tremor, a dopaminergic deficit is present [67]. Dissociation of parkinsonian tremor and hypokinetic symptoms may be due to the different patterns of nigral degeneration [6]. Thus, the variations in the dopamine level are likely to act in the way in which they are considered in the model. Moreover, if the effect of cholinergic or other pathologies in Parkinson's disease is to increase effective coupling in the basal ganglia-thalamo-cortical circuitry, these pathologies are likely to induce tremor-like bursting. This is expected because our modeling indicates that the lack of dopamine promotes oscillations due to the increase in the coupling in the circuits.

Compensatory effects are not considered in our model although they have been conjectured to play a role in the tremor genesis (e.g., [7]). Compensatory effects may slow down the increase of the feedback strength or may even eventually weaken it as a result of overcompensation (which may be one of the explanations for why the tremor severity may decrease in the advanced state of Parkinson's disease). However, the variation of feedback strength is not removed by these kinds of compensations, rather the timing of the processes and its magnitude are altered. Therefore the modeling conclusions are unlikely to be invalidated by the presence of compensation.

While this paper studies the network and cellular properties affecting the tremor-like oscillations, there is an important question about interpretation of the mechanisms of this phenomenon in more mathematical terms using dynamical systems theory. This appears to be a challenging problem as delays introduce an infinite-dimensional dynamical system. Moreover, the time scale of spiking is only a few times larger than the time scale of bursting, which may potentially make the separation of time scales and resultant perturbative analysis difficult.

Finally, it is known, that the neural activity in tremor-supporting networks exhibits a complex spatio-temporal structure [68]. These patterns are likely to be induced by the complex anatomical structure of the tremor-supporting networks and thus cannot be reproduced in our model.

*Implications for the tremor-genesis and tremor therapies*
Earlier indirect evidences (discussed in Introduction) suggested that parkinsonian tremor arises in the basal ganglia-thalamo-cortical loops, and that the presence of the thalamocortical feedback to basal ganglia is essential for tremor occurrence. However, there was no direct experimental study of this hypothesis. Such a study is clearly hard to implement. *In vitro* preparations will not be able to maintain the structure of the loop which spans multiple subcortical and cortical locations. *In vivo* studies would be limited by the difficulty of recording from multiple locations of the circuitry and with variation of multiple parameters. Available animal models of Parkinson's disease either do not exhibit tremor at all or exhibit tremor, which is not really similar to the human parkinsonian tremor [69,70]. In these circumstances, the computational neuroscience approaches become especially valuable.

The minimalistic representation of the thalamo-cortical feedback in the present modeling study signifies a very general role of this feedback in the tremor genesis. This study suggests that just the presence of the relatively strong basal ganglia-thalamo-cortical feedback leads to the birth of tremor-like oscillations under rather general conditions. The study indicates that the parkinsonian tremor genesis has its origin in both the properties of local basal ganglia circuits and in the thalamo-cortical feedback to the basal ganglia. This feedback loop is modulated by dopamine and as dopamine level decreases, the strength of the loop increases to generate tremor-like oscillations. While weakly-connected (presumably normal) cortico-subcortical loops through the basal ganglia and thalamus may be crucial for movement control and other functions, we show that malfunctioning of modulatory mechanisms of these loops can cause tremor.

The feedback-loop mediated origin of tremor suggests some new directions for its treatment. Since the weakening or destruction of the loop itself (rather than a particular node of the loop network) may suppress tremor, different sites in the loop network may be explored as anatomical targets for surgical or pharmacological intervention. Lesions or high-frequency stimulation beyond traditional pallidal or thalamic sites may turn out to be efficient. This prediction is testable. Trying surgical ablation or stimulation outside of basal ganglia may be risky for a variety of factors (and, as we explained above, is unlikely to be relevant in animal models). However, a potential confirmation for this prediction may come from disruption of cortical activity through non-invasive transcranial magnetic stimulation. Similarly, pharmacological influences of different nature may prove to be effective as long as they appropriately decrease the strength of the basal ganglia-thalamo-cortical loop at any of its parts. As the intracranial drug delivery targeted to a specific location may become more feasible in patients, targeting a delivery of inhibitory modulator to some parts of cortico-basal ganglia-thalamocortical loop may be effective too.

**Acknowledgements**
We thank Dr. K.A. Sigvardt for discussions of the previous version of the model, Dr. D. Terman for sharing his model xpp files, Dr. C. Park for assistance with computations, and Dr. R.M. Worth for his comments on the manuscript. This study was supported by the National Institutes of Health grant R01NS067200 (NSF/NIH CRCNS program).